\newcommand{\black}[1]{\textcolor{black}{#1}}

\documentclass[trackchanges]{aastex701}
\begin{document}

\title{Discovery of Short-Term $\gamma$-Ray Pulsed Radiation Variations Following a Glitch in PSR J0205+6449}

\author[orcid=0000-0002-1662-7735]{Wen-Tao Ye}
\affiliation{State Key Laboratory of Particle Astrophysics, Institute of High Energy Physics, Chinese Academy of Sciences, Beijing 100049, People’s Republic of China}
\affiliation{University of Chinese Academy of Sciences, Chinese Academy of Sciences, Beijing 100049, People’s Republic of China}
\email[show]{yewt@ihep.ac.cn}

\author[]{Ming-Yu Ge}
\affiliation{State Key Laboratory of Particle Astrophysics, Institute of High Energy Physics, Chinese Academy of Sciences, Beijing 100049, People’s Republic of China}
\email[show]{gemy@ihep.ac.cn}  

\author[]{Shi-Jie Zheng}
\affiliation{State Key Laboratory of Particle Astrophysics, Institute of High Energy Physics, Chinese Academy of Sciences, Beijing 100049, People’s Republic of China}
\email[]{zhengsj@ihep.ac.cn} 

\author[]{Xiang Ma}
\affiliation{State Key Laboratory of Particle Astrophysics, Institute of High Energy Physics, Chinese Academy of Sciences, Beijing 100049, People’s Republic of China}
\email[]{}

\author[]{Yu-Jia Zheng}
\affiliation{Department of Physics and Institute of Theoretical Physics, Nanjing Normal University, Nanjing, 210023, Jiangsu, China}
\affiliation{State Key Laboratory of Particle Astrophysics, Institute of High Energy Physics, Chinese Academy of Sciences, Beijing 100049, People’s Republic of China}
\email[]{}

\author[]{Han-Long Peng}
\affiliation{Department of Physics and Institute of Theoretical Physics, Nanjing Normal University, Nanjing, 210023, Jiangsu, China}
\affiliation{State Key Laboratory of Particle Astrophysics, Institute of High Energy Physics, Chinese Academy of Sciences, Beijing 100049, People’s Republic of China}
\email[]{} 

\author[]{Xue-Zhi Liu}
\affiliation{Institute of Astrophysics, Central China Normal University, Wuhan 430079, China}
\affiliation{State Key Laboratory of Particle Astrophysics, Institute of High Energy Physics, Chinese Academy of Sciences, Beijing 100049, People’s Republic of China}
\email[]{}

\author[]{Li-Ming Song}
\affiliation{State Key Laboratory of Particle Astrophysics, Institute of High Energy Physics, Chinese Academy of Sciences, Beijing 100049, People’s Republic of China}
\email[]{songlm@ihep.ac.cn} 

\author[]{Shuang-Nan Zhang}
\affiliation{State Key Laboratory of Particle Astrophysics, Institute of High Energy Physics, Chinese Academy of Sciences, Beijing 100049, People’s Republic of China}
\email[]{zhangsn@ihep.ac.cn}

\author[]{Fang-Jun Lu}
\affiliation{State Key Laboratory of Particle Astrophysics, Institute of High Energy Physics, Chinese Academy of Sciences, Beijing 100049, People’s Republic of China}
\email[show]{lufj@ihep.ac.cn}

\begin{abstract}
Rotation-powered pulsars exhibit stable emission characteristics most of the time. However, their radiative state can vary with the sudden changes of rotational state such as glitches. To date, pulsed radiation changes associated with glitches have only been detected in the radio band. Since the emission regions of radio and $\gamma$-ray may differ, searching and investigating whether glitches can induce changes in high-energy radiation would further deepen our understanding of how glitches affect the magnetosphere of pulsars. We report successive variations in the $\gamma$-ray pulsed radiation of PSR J0205+6449 following the glitch at MJD 54904 observed by the {\sl Fermi}/LAT. The amplitude ratio of the two peaks showed a hint of an increase during MJD 54905--54940 initially, followed by a recovery to the mean level and a significant ($>5\,\sigma$) decrease in the separation between the two peaks over MJD 54940--55000. The amplitude ratio of the two peaks increased ($\sim3\,\sigma$) again in MJD 55000--55160, accompanied by a marginal flux variation. Finally, the pulsed radiation reverted to its normal state. This is the first significant detection of pulsed radiation variation associated with a glitch in $\gamma$-ray pulsars. We attribute this to magnetospheric reconfiguration triggered by localized crustal breaking and associated elastic displacement near the polar cap following the glitch.

\end{abstract}


\section{Introduction}

Rotation-powered pulsars generally have stable emission features. However, accompanied by changes in the rotational state including spin frequency $\nu$ and spin-down rate $\dot\nu$, the radiation state may also change. 

Two typical types of rotational changes are recognized. One is the spin-down state change, in which the spin-down rate $\dot\nu$ changes suddenly and persistently, while the spin frequency $\nu$ remains continuous and often exhibits quasi-periodicity \citep{1971_mode_change,Lyne_2010_mode_switch}. Such phenomena have abundant observational samples in the radio band and are often accompanied by pulse profile changes \citep{2018_kouff_state_change,2022_17pulsars_state_change}. In the high-energy band, state changes have also been detected in PSR B0540+69, PSR J1124-5916, and PSR J2021+4026 \citep{PSRJ2021+4026,b0540_mode_change,J1124-5916state_transition}. PSR J2021+4026 is the first known variable $\gamma$-ray pulsar: following a state change, its pulse profile changed from triple-peaked to double-peaked, and its flux above 100\,MeV decreased by 20\%. It exhibits repeated behavior similar to that observed in radio pulsar samples \citep{repeated_j2021,2024_repeated_j2021}. Such state changes are believed to be caused by the global reconfiguration of the pulsar’s magnetosphere \citep{2010_nulling_model,state_change_model_interpretation,J2021_simulation}.

The other type is the glitch, during which both the pulsar's spin frequency $\nu$ and spin-down rate $\dot{\nu}$ undergo sudden changes. Glitches are thought to be caused by the transfer of angular momentum from the inner rapidly rotating component to the outer crust \citep{1975_glitch,1984_glitch,2012_glitch,2014_glitch,2022_gltich_review}. They are commonly observed in both radio and high-energy pulsars. Thanks to the high sensitivity of radio telescopes, radiation variations following glitches have been detected in several pulsars, including the direct observation of the missing of single pulse during a \black{glitch} in the Vela pulsar \citep{velaemissionchange,radio_J2021+51,radio_B1822-09_flux_tube,radioJ0738-4042,2024_1048_liu,2025_radio_pulse_change_0742}. 
In contrast, the high-energy pulse profiles and fluxes of isolated pulsars such as the Crab and Vela pulsars remain stable after glitches \citep{2020_crab_correlation,crabnovary,vela}. Several studies have attempted to search for pulsed radiation variations associated with glitches, but no definitive correlation \black{between glitch events and high-energy radiative changes has been identified to date \citep{PSRJ1420-6048,2024_1048_liu,2025_search_gamma_vary,2025_searching2,2025_J1522}}. This null result may be attributed to the long timescales required to accumulate sufficient photons for a well pulse profile in high-energy band, such integration times can smooth out short-term variability. Consequently, there is currently no evidence that glitches in rotation-powered pulsars induce detectable variations in their high-energy pulsed radiation.

High-energy radiation from rotation-powered pulsars is believed to originate not from the polar cap, but from the gap region near the null charge surface \citep{cks1986,harding_1994,out_gap_romani_1996,twopolemodel,qiao2004}. \black{Moreover, recent PIC and MHD simulations indicate that the dominant high-energy $\gamma$-ray radiation originates in the current sheet near the Y-point\citep{2014_current_Kalapotharakos,2014ApJ_Philippov,2015ApJ_Philippov,2018ApJ_Brambilla,2022_Philippov}.} Variations in the pulsed radiation can reflect changes in the emission region and thus reveal magnetospheric alterations. Therefore, searching for and studying the impact of glitches on the high-energy radiative state of rotation-powered pulsars is of great significance for understanding how glitches affect the magnetosphere.

Young high-energy pulsars exhibit frequent, large glitches and high spin-down luminosities, making them ideal targets for investigating the relationship between pulsed radiation and glitches. Among them, PSR J0205+6449 is a notable source due to its frequent glitches and bright $\gamma$-ray emission. It is associated with the supernova remnant 3C58 and was first discovered via its X-ray pulsations by the {\sl Chandra} Telescope \citep{Chandra_discover_2002}. Pulsed emission was subsequently detected in the radio band by the Green Bank Telescope \citep{radio_obser_2022} and in $\gamma$-ray by the {\sl Fermi} Gamma-ray Space Telescope \citep{fermi_obser_2009}. This pulsar is located at a distance of approximately 2 kpc from Earth \citep{0205_distance}. The spin frequency $\nu\sim15.2\ \rm{Hz}$ and spin-down rate $\dot\nu\sim4.5\times10^{-11}\ \rm{Hz}\ s^{-1}$ of PSR J0205+6449 yield a dipole magnetic field of
$3.6\times{10^{12}}\,{\rm G}$, a characteristic age of 5400 years, and a very high spin-down luminosity of $2.7\times{10^{37}}\,{\rm erg\,s^{-1}}$, making it the third most energetic of the known Galactic pulsars \citep{Chandra_discover_2002}. 

The pulse profile of PSR J0205+6449 exhibits a double-peaked structure in the X-ray and $\gamma$-ray bands with a peak separation of 0.49 in phase, but appears single-peaked in the radio band \citep{radio_obser_2022,0205flux_ratio_change,0205gammalocation}. \cite{0205flux_ratio_change} analyzed the {\sl RXTE} observations of PSR J0205+6449 and found that the flux ratio of the two pulse peaks deviated by $3.5\,\sigma$ from the mean in one epoch, with no accompanying change in the pulse profile. This anomaly suggests that PSR J0205+6449 may exhibit more intriguing phenomena that require further investigation.

In this work, we report successive changes in the $\gamma$-ray pulsed radiation of PSR J0205+6449 following a glitch at MJD 54904 detected by {\sl Fermi}/LAT. In Section \ref{method}, the {\sl Fermi}/LAT data selection and analysis methods are described. In Section \ref{result}, the timing results, as well as variations in the pulse profile and flux are presented. Finally, in Section \ref{discussion}, we discuss how glitches affect the magnetosphere of the pulsar.

\section{Data and Analysis}\label{method}

The {\sl Fermi}/LAT is a $\gamma$-ray telescope with a wide field of view, large effective area, and an energy range of 20 MeV to 300\,GeV \citep{FERMIlat}. Since 2008 August 3, the LAT has been operating in survey mode, completing a full-sky scan approximately every three hours. This continuous observation capability provides a unique opportunity to study the spin evolution and high-energy emission variability of $\gamma$-ray pulsars.

The Pass 8 {\sl Fermi}/LAT event data of PSR J0205+6449 spanning from MJD 54700 to 60000 are selected and processed using Fermitools (version 2.2.0) \citep{fermianalysis}. \black{For likelihood analysis,} photon events in the energy range of 0.1--10\,GeV, with \texttt{evclass}=128, \texttt{evtype}=3, a circular Region of Interest (RoI) of radius $18^{\circ}$, and a zenith angle $<90^{\circ}$, are selected using \texttt{gtselect}.  The arrival times of all events are corrected to the solar system barycenter using \texttt{gtbary}, adopting the solar system ephemeris DE405 and the pulsar position $\rm{RA=02^h05^m37.9^s}$ and $\rm{DEC=64^{\circ}49^{'}42.8^{''}}$ \citep{2013_vlbi,0205gammalocation}.

\black{PSR J0205+6449 lies in the anti-Galactic center direction at Galactic coordinates $(l, b) = (130.7^\circ, 3.1^\circ)$. The region exhibits low levels of diffuse emission and is sparsely populated by bright neighboring sources \citep{2012_fermi_Diffuse,0205gammalocation}. To minimize background contamination in pulsar timing and pulse profile analyses, we further select photon events in the energy range 0.35--10\,GeV from PSR J0205+6449}, and apply an energy-dependent acceptance cone $\Theta_{1\sigma}(E)$ that encloses approximately 68\% of the point-spread function (PSF) in each 10\,MeV energy bin, as done in \cite{kuiper2018}. \black{The typical PSF is $1.3^{\circ}$ at 0.35\,GeV and $0.6^{\circ}$ at 1\,GeV}. This selection effectively suppresses the background and enhances the significance of the pulse profile.

We determined the rotational frequency $\nu$ and its first derivative $\dot{\nu}$ by maximizing the $\chi^2$ search in 60-day intervals. Subsequently, we folded the pulse profiles accumulated over 10-day exposure periods to derive the pulse times of arrival (TOAs). Phase-coherent timing analysis is implemented using the \texttt{Tempo2} software package to fit the rotational parameters. PSR J0205+6449 exhibits frequent glitches and significant timing noise \citep{radio_xray_timming_noise,0205flux_ratio_change}. To mitigate timing noise and accurately constrain glitch occurrence times, we iteratively refined the rotational ephemeris. In each iteration, we used the current ephemeris to re-derive the pulse times of arrival (TOAs), then fitted the timing model to these TOAs to update the ephemeris. This process continued until the phase residuals were consistent with the model in Equation (\ref{eq1}): 
\begin{equation}
\phi ( t ) = \phi _ { 0 } + \nu ( t - t _ { 0 } ) + \frac { 1 } { 2 } \dot { \nu } ( t - t _ { 0 } ) ^ { 2 } + \frac { 1 } { 6 } \ddot { \nu } ( t - t _ { 0 } ) ^ { 3 } ... ,    
\label{eq1}
\end{equation} 
where $\nu$, $\dot{\nu}$ and $\ddot { \nu }$ are the spin frequency and its first and second time derivatives at the reference epoch $t_0$. As for glitch, the post-glitch rotational parameters can be described by Equation (\ref{eq:glitch_phi}) \citep{1992_lyne,2020_crab_delay_spin_up}:

\begin{equation}
\phi_{\rm{g}}(t) =\phi ( t )+ \Delta\nu_{\rm{p}} \Delta t + \frac{1}{2} \Delta\dot{\nu}_{\rm{p}} \Delta t^2
+ \Delta\nu_{\rm{d}} \tau_{\rm{}} \left(1 - \exp\left(-\frac{\Delta t}{\tau_{\rm{}}}\right)\right),
\label{eq:glitch_phi}
\end{equation}
where $\Delta t = t - t_{\rm g}$ is the time after the glitch epoch $t_{\rm g}$, $\Delta \nu_{\rm p}$ and $\Delta \dot{\nu}_{\rm p}$ are the
permanent changes of $\nu$ and $\dot{\nu}$, $\tau$ is the time scale of delayed spin-up
process.

For the analysis of profile variations, the 15-year observational data of PSR J0205+6449 are processed via partial-timing, \black{which involves fitting for ephemerides over short intervals to mitigate the effects of long-term timing noise and to calculate the arrival phase of each photon}. The template pulse profile is generated by folding all pulse phases. The short-term pulse profiles are extracted from this phase dataset for the targeted time periods. We define the primary peak near phase 0.9 as P1 and the secondary peak near phase 0.4 as P2. Phase 0 is set at the trailing edge of P1. For the comparative analysis between the short-term profiles and the template profile, we adopt P1 to align the pulse phases and normalize the peak amplitudes. The pulse profiles are normalized as $N_{\rm{norm}}=(N-N_{\rm{bkg}})/N_{\rm{p1}}$, where $N$ is the number of counts of each phase bin, $N_{\rm{bkg}}$ is the mean counts in off-pulse phases of 0--0.3 and 0.5--0.7, and the $N_{\rm{p1}}$ is the mean counts of P1 in the phase range of 0.7-1. \black{Subsequently, the template profile is rescaled so that the peak of its P1 component equals 1, and all short-term profiles are scaled by the identical factor.} The significance of the discrepancy between the two pulse profiles is estimated by the $\chi^2$ statistic calculated from their residuals. Gaussian functions are used to fit the pulse profile to quantify the variations in the amplitude and phase of the pulse peaks.

We perform a \black{binned} likelihood analysis for flux estimation using \texttt{Fermipy} (version 1.2.0) \citep{fermipy}, which is built on \texttt{Fermitools} (version 2.2.0). The spectral model includes all sources from the 4FGL-DR4 catalog \citep{4th_cata} within a $15^{\circ}$ region centered on the target source, together with the Galactic diffuse emission model \texttt{gll\_iem\_v07.fits} and the isotropic diffuse background model \texttt{iso\_P8R3\_SOURCE\_V3\_v1.txt}. We first free all sources within $5^{\circ}$ to fit the global spectral model. Subsequently, only the normalizations of the diffuse background components and sources within $3^{\circ}$ are freed to fit the short-term flux of PSR J0205+6449. \black{Since the PWN emission is negligible compared to the pulsar emission below 10 GeV, fitting with two spectral components is not required for short-timescale variability studies \citep{0205gammalocation}.}

\section{result}\label{result}
\subsection{Timing Results}

\begin{figure}[h]
\centering

\includegraphics[width=1\linewidth]{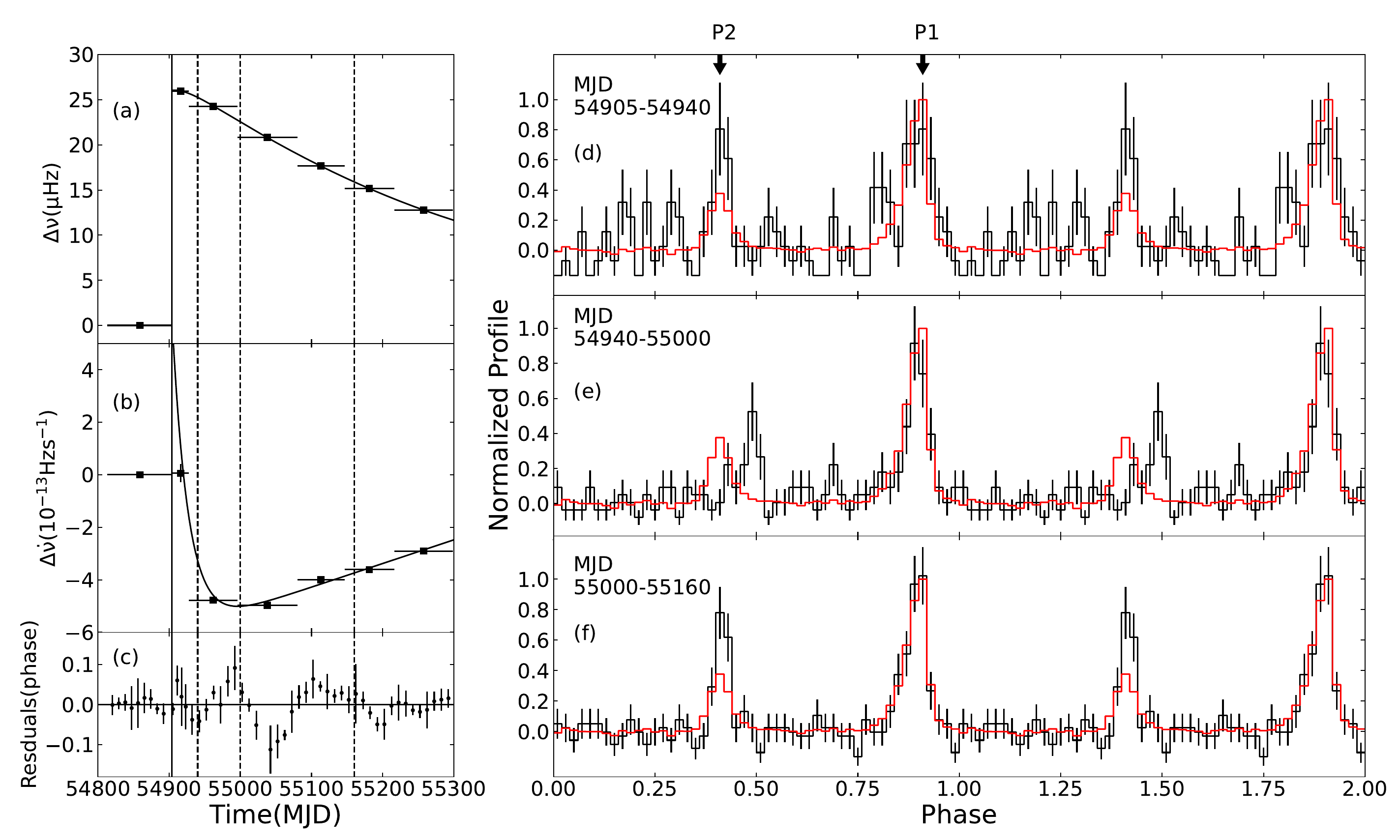}
\caption{The left panel shows the timing results for the glitch at MJD 54904. Panel (a) and (b) show the residuals of $\nu$ and $\dot{\nu}$ of partial-timing results that subtract the pre-glitch spin-down trend, respectively. The solid lines represent the glitch model listed in Table \ref{tar:timingpar}. Panel (c) displays the TOA residuals fitted with the glitch model listed in Table \ref{tar:timingpar}. The right panel shows short-term variations in the pulse profile, and the vertical dashed lines in left panel mark their epochs. Panels (d), (e), and (f) show the three successive pulse profile variations during MJD 54905--54940, MJD 54940--55000, and MJD 55000--55160, respectively. The black profile represents the short-term pulse profile, and the red profile represents the template profile accumulated from full data.}
\label{fig:timing_result}
\end{figure}

The glitches of PSR J0205+6449 have been reported in \citep{glitch_detection,2024_0205_timing}. The glitch investigated in this work occurred at 54907(18), with the measured changes in $\Delta\nu/\nu$ and the $\Delta\dot{\nu}/\dot{\nu}$ are $1761(7)\times10^{-9}$ and $12(2)\times10^{-3}$, respectively. Benefiting from the continuous monitoring by \textit{Fermi}/LAT, our detailed analysis reveals a negative exponential recovery following the glitch at MJD 54904, suggesting that this event is a delayed spin-up glitch.
The rotational state over a 400-day span after this glitch can be described by equation (\ref{eq:glitch_phi}), with the parameters listed in Table \ref{tar:timingpar}. The model-predicted $\nu$ and $\dot{\nu}$ are shown as the solid lines in panels (a) and (b) of Figure \ref{fig:timing_result}, respectively, with the data points representing the partial-timing measurements. The delayed spin-up process is evident in the evolution of $\dot{\nu}$. The timing residuals displayed in panel (c) of Figure \ref{fig:timing_result}, exhibit significant timing noise. To obtain accurate short-term pulse profiles, we divide the 300-day interval following the glitch into three sub-intervals and derive an independent rotational ephemeris for each. The resulting ephemerides are listed in Table \ref{tar:timingpar}. Our partial-timing analysis of the delayed spin-up recovery yields a decay timescale of $\sim9$ days. This value is significantly longer than those observed in the Crab pulsar, which range from 0.6 to 2.6 days \citep{2020_crab_delay_spin_up}, but is close to the decay timescale of $\sim8$ days reported for the magnetar SGR J1935+2154 \citep{2024_ge_1935}.

\begin{table}[htbp]  
    \centering
    \caption{Timing results of PSR J0205+6449 around the glitch at MJD 54904}
    \begin{tabular}{lc}%{0.3\textwidth}

        \hline
        Parameter & Value\\
        
        \hline
        $\rm{Glitch\ Timing\ Results}$ \\
        
        \hline
        $\rm{Time\  Range(MJD)}$ & 54800-55300  \\
        $ \rm{Epoch(MJD)}$ & 54904 \\
        $ \rm{Glitch\,Epoch(MJD)}$ & 54904(5) \\
        $\nu\rm{(Hz)}$ & 15.213938150(6) \\
        $\dot{\nu}\rm{(10^{-11}\,Hz\ s^{-1})}$ & --4.4730(1)\\
        $\black{\ddot{\nu}\rm{(10^{-21}\,Hz\ s^{-2})}}$ & 1.034(9)\\
        $\Delta\nu_{p}\rm{(\mu Hz)}$ & 21.29(8) \\
        $\Delta\dot{\nu_{p}}\rm{(10^{-13}\,Hz\ s^{-1})}$ & 2.18(3)\\
        $\rm{\nu_d(\mu hz)} $ & -1.54(4)\\
        $\rm{\tau(days)} $ & 20(1) \\

        \hline
        $\rm{Partial-Timing\ Results}$ \\
        
        \hline
        $\rm{Time\  Range(MJD)}$ & 54900-55040  \\
        $\rm{Epoch(MJD)}$ & 54915 \\
        $ \rm{Glitch\,Epoch(MJD)}$ & 54904 \\
        $\nu\rm{(Hz)}$ & 15.21393743(1) \\
        $\dot{\nu}\rm{(10^{-11}\,Hz\ s^{-1})}$ & -4.5274(2)  \\
        $\rm{\nu_d(\mu Hz)} $ & -1.17(8)\\
        $\rm{\tau(days)} $ & 8.8(1) \\
        $\rm{\phi_{0}}$ & -0.454 \\
        $\chi^2 / dof$ & 4/9 \\
        
        \hline
        
        $\rm{Time\  Range(MJD)}$ & 55000-55080  \\
        $\rm{Epoch(MJD)}$ & 55070.2 \\
        $\nu\rm{(Hz)}$ & 15.213330272(6) \\
        $\dot{\nu}\rm{(10^{-11}\,Hz\ s^{-1})}$ & -4.5284(2)  \\
        $\rm{\phi_{0}}$ & -0.236 \\
        $\chi^2 / \rm{dof}$ & 4/6 \\
        \hline
        
        $\rm{Time\  Range(MJD)}$ & 55080-55200  \\
        $\rm{Epoch(MJD)}$ &  55140.2 \\
        $\nu\rm{(Hz)}$ & 15.213056902(2) \\
        $\dot{\nu}\rm{(10^{-11}\,Hz\ s^{-1})}$ & -4.51752(8)  \\
        $\black{\ddot{\nu}\rm{(10^{-21}\,Hz\ s^{-2})}}$ & 10.1(1) \\
        $\rm{\phi_{0}}$ & 0.0842 \\
        $\chi^2 / \rm{dof}$ & 8/10\\
        \hline

        \end{tabular}
    \label{tar:timingpar}
\end{table}

\subsection{Pulse Profile Variations}

\begin{figure}[h]
\centering
\includegraphics[width=0.5\linewidth]{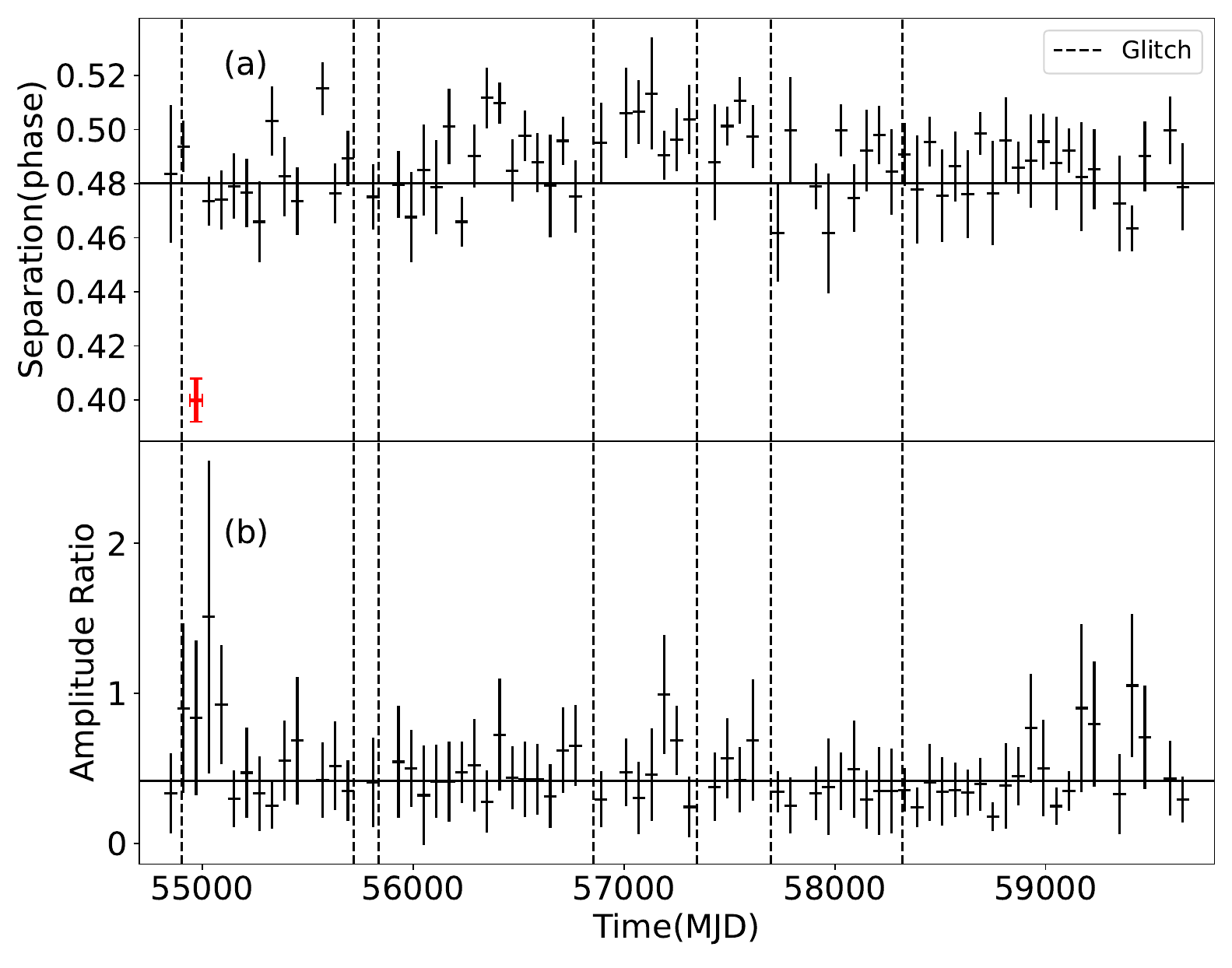}
\caption{Pulse peak separation and amplitude ratio measured in 60-day intervals. Panel (a) shows the peak separation, and panel (b) shows the amplitude ratio. Vertical dashed lines mark glitch epochs. A significant anomaly (highlighted in red) appears near the glitch at MJD 54904, with no notable variations detected at other glitch epochs or during other intervals.}
\label{fig:pa_evolution}
\end{figure}

Following the glitch, we identify three consecutive variations in the pulse profile, corresponding to the time intervals MJD 54905--54940, MJD 54940--55000, and MJD 55000--55160. The short-term pulse profiles are folded by the partial-timing ephemerides listed in Table \ref{tar:timingpar}. The three short-term pulse profiles are normalized and plotted in Figure \ref{fig:timing_result} (d), (e) and (f), respectively. Within approximately 35 days after the glitch, \black{the Gaussian fitting results show that the relative amplitude of P2 increased by 153(94)\%, with a significance level of $\sim1.5\,\sigma\ (\chi^2/\rm{dof}=4/2)$}. During the 60-day interval from MJD 54940 to 55000, the amplitude of P2 returned to its normal level, and the separation between the two peaks decreased by 0.075(8) in phase, with the discrepancy relative to the template profile reaching a significance level of $>5\,\sigma\ (\chi^2/\rm{dof}=56/7)$. Subsequently, over the 160-day interval from MJD 55000 to 55160, the peak separation recovered, while the amplitude of P2 increased again by 148(73)\%, with a significance level of $\sim3\,\sigma\ (\chi^2/\rm{dof}=11/2)$. Thereafter, the pulse profile reverted to its normal state, showing no identifiable phase shift or peak amplitude variation. We also examine the variations in pulse peak separation and amplitude ratio (P2/P1) in 60-day intervals across the full 15-year dataset, with the results shown in Figure \ref{fig:pa_evolution}. No significant variations are observed at other times or in association with other glitches. \black{Since both pulse peaks are narrow in phase, reliable measurement of changes in pulse profile shape parameters (e.g., width) is not feasible with the current data quality.}

\subsection{Flux Variations}

\begin{figure}[h]
\centering

\includegraphics[width=0.5\linewidth]{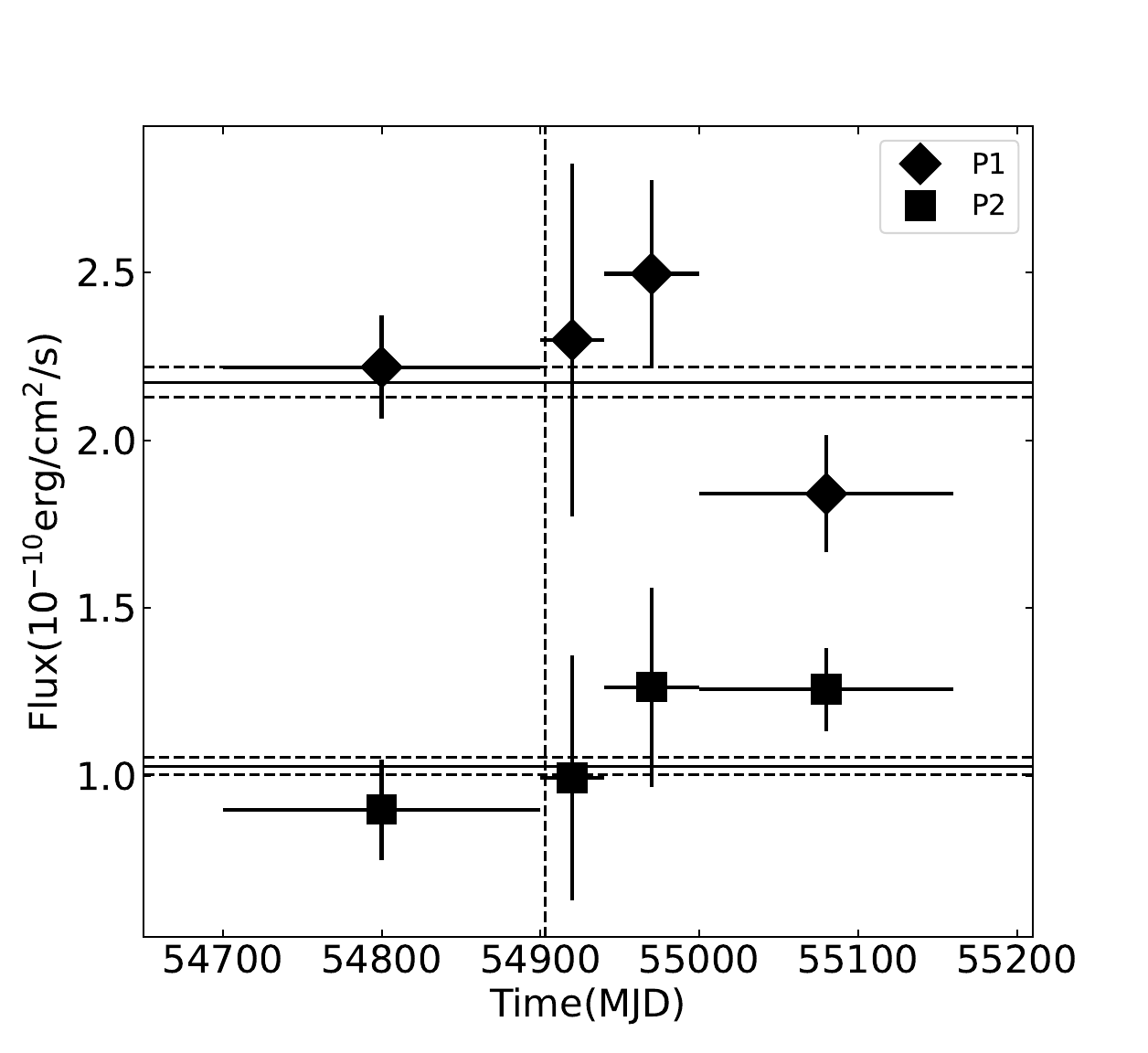}
\caption{The flux of the two pulse peaks. The diamond and square points represent the flux of P1 and P2, respectively. The black vertical dashed lines represent the occur time of glitch. The horizontal solid line at 2.17 and 1.02 are the mean flux of P1 and P2, and the dashed lines represent their 1$\,\sigma$ range.
\label{fig:flux}}
\end{figure}
We perform flux analysis for the two pulse peaks using the phase ranges 0.36--0.5 for P1 and 0.8--0.96 for P2. To maintain consistency with the pulse profile analysis, the same time intervals are used for flux estimation. As shown in Figure \ref{fig:flux}, the solid lines at 2.17 and 1.02 (in units of $10^{-10}\,\mathrm{erg/cm^{2}/s}$) represent the long-term average fluxes of P1 and P2, respectively. Due to the low count rates and integration intervals, no significant flux variations are detected during MJD 54905--54940 and MJD 54940--55000. However, in MJD 55000--55160, both peaks exhibit a marginal flux variation: the flux of P1 decreased of 22 (12)\%, while the flux of P2 increased of 18 (10)\%. These results suggest that the observed increase in the peak amplitude ratio during this interval may be attribute not only to an increase in P2, but also to a concurrent decrease in P1.

\black{To test for corresponding spectral variations, we also perform a binned likelihood analysis with free spectral parameters. However, due to limited photon statistics, we do not find robust evidence for short-term spectral variability.}

\section{Discussion}\label{discussion}

Radio observations of the Vela pulsar have shown that its emission mode undergoes rapid alternation on very short timescales during a glitch. This provides direct evidence that glitches can influence the magnetosphere \citep{velaemissionchange}. In other radio studies of pulse profile changes associated with glitches, the observed variations are generally attributed to glitch-induced displacements of magnetic flux tubes near the emission regions \citep{radio_J2021+51,radio_B1822-09_flux_tube,2024_1048_liu,2025_radio_pulse_change_0742}. A theoretical explanation for how a glitch drives such displacements based on the vortex creep model was presented in a study of PSR J0738-4042 \citep{vortex_flux_tube,erbil_2022,radioJ0738-4042}. That study also points out that these changes are hard to be observable in high-energy pulse profiles, as radio and $\gamma$-ray emissions originate from distinct regions of the magnetosphere.

A few studies have predicted that glitches can cause variations in high-energy radiation. For example, stellar quakes may induce crustal plate motions that redistribute the magnetic field and alter the magnetic inclination angle, thereby modifying the observed pulse profile \citep{quaking_pulsar}. Moreover, simulations of $\gamma$-ray pulse profiles show that magnetic field perturbations can lead to changes in both the peak phase and peak intensity of the pulse profile \citep{gamma_pulsar_simulation,self_consistent_electric,self_consistent_electric_2}. Therefore, numerous studies have attempted to detect variable pulsed radiation associated with glitches in bright pulsars such as the Crab and Vela, yet no evidence has been found that glitches can affect high-energy radiation \citep{2020_crab_correlation,vela,2025_search_gamma_vary}. Even for the largest glitch, which was also a delayed spin-up event observed in the Crab pulsar at MJD 58064, the analysis of high-SNR X-ray data revealed no detectable changes in the pulse profile or flux \citep{crab_polorization_vary,2020_crab_delay_spin_up,crabnovary}. This indicates that neither the type nor the magnitude of a glitch is the primary factor responsible for changes in pulsed radiation. 

\black{Pulsar timing results show that PSR J0205+6449 exhibited a change in its spin frequency derivative ($\dot{\nu}$) of less than 0.1\,\% between two episodes of radiation variation around MJD 55000, far smaller than the approximately 3\,\% variation in $\dot{\nu}$ observed in PSR J2021+4026 during its state transition \citep{PSRJ2021+4026}. Moreover, $\dot{\nu}$ evolved continuously throughout the transitions between different emission states, indicating that the magnetosphere of PSR J0205+6449 did not undergo a global reconfiguration and thus did not significantly affect its magnetic braking torque.} Notably, for PSR J0205+6449, we have detected only one pulsed radiation variation event following the glitch at MJD 54904, and no similar variation is observed after several other glitches, \black{even those with larger changes in spin frequency $\nu$  and its first derivative $\dot{\nu}$. Therefore, there is no indication that the radiation variation in PSR J0205+6449 exhibits a strong correlation with the spin-down rate variation, similar to PSR J2021+4026 \citep{2026_2021_liu}.}

We suggest that the sporadic nature of pulsed radiation variations associated with glitches may arise from the spatially localized origin of glitches in the pulsar crust. The location where a glitch occurs is the crucial factor determining whether the glitch can affect the magnetosphere. In this event, the crustal region perturbed by the glitch was situated near the polar cap, where the footpoints of magnetic field lines are concentrated. Consequently, localized fracture and elastic displacement in the crust perturbed the magnetic field lines anchored  near the polar cap, inducing changes in the magnetospheric structure. \black{This triggered a reconfiguration of the magnetosphere, which could alter current flow, particle injection, and conductivity, thereby impacting particle acceleration, including the geometry of acceleration zones and the beaming direction, as revealed by PIC simulations \citep{2018_Philippov_emissionmodel,2022_Philippov}. This, in turn, eventually led to a series of unpredictable variations in the observed radiation properties over hundreds of days.}

Multi-wavelength coordinated observations of such events will yield richer scientific insights. For instance, because the pulsar’s magnetic field may contain significant multipolar components near the surface and X-ray emission originates from magnetospheric gap regions (e.g., outer gap) \citep{cks1986,2024_repeated_j2021}, variations in pulse radiation may be more readily detectable in the X-ray band. \black{Specifically, if the altitude or location of the accelerator changes, corresponding variations in non-thermal pulse radiation could be detected in the X-ray and $\gamma$-ray bands.} If post-glitch variations of pulsed emission are simultaneously observed in both $\gamma$-ray and X-ray bands, examining the differences in how the emission states vary across energy bands will impose stronger constraints on magnetospheric structure and high-energy emission mechanisms. \black{Additionally, high-energy particles flowing back along the magnetic field lines heat the stellar surface, forming hot spots in the polar cap region where pulsed thermal radiation can be detected \citep{2022hot_point}. If a glitch induces a change in the surface magnetic field, the hot spots will shift position. This displacement can be detected in the soft X-ray pulse profiles and phase-resolved spectra. Furthermore, Radio emission is thought to be beamed along the magnetic axis and is more sensitive to variations, whereas high-energy radiation originates from higher regions of the magnetosphere and is relatively less susceptible to disturbances \citep{1975_polar_gaps,radioJ0738-4042}. Therefore}, conducting simultaneous pulsar observations in radio and high-energy bands around the time of a glitch and analyzing the phase alignment between the two bands before and after the event can reveal whether the magnetic inclination angle changed after the glitch \citep{2013_b1943,2016_B0611}. \black{Moreover, if large-scale magnetic field reconfiguration occurs, the pulse profiles in both radio and high-energy bands will change simultaneously. In this scenario, modeling the emission in each band before and after the glitch is required. Subsequently, tracking the evolution of these pulse profiles and the phase lag between the two bands can provide further insights into systematic changes in the magnetosphere and the magnetic inclination angle.}

Overall, changes in high-energy radiation associated with glitches may be common. More similar phenomena remain to be identified. In the future, high-sensitivity time-domain telescopes such as eXTP are expected to detect additional such events in the high-energy band \citep{2025_extp,2025_extp_mageta}, facilitating a better characterization of variability processes and a deeper understanding of the underlying physics.

\section{summary}

We report successive changes in the $\gamma$-ray pulsed radiation of PSR J0205+6449 following the glitch at MJD 54904. The amplitude ratio of the two peaks initially increased, followed by a return to the mean level and a decrease in their phase separation. Subsequently, the amplitude ratio increased again, accompanied by a marginal change in flux. These changes occurred on timescales of approximately 35, 60, and 160 days, respectively. The pulsed radiation then returned to its quiescent state. This is the first significant change in $\gamma$-ray pulsed radiation associated with a glitch, demonstrating that glitches can perturb the magnetosphere of rotation-powered pulsars and alter their high-energy radiation on timescales of hundreds of days. We attribute this behavior to magnetospheric reconfiguration triggered by a glitch that occurred near the polar cap.

\section*{Acknowledgement}

This research utilized data and software from the High Energy Astrophysics Science Archive Research Center (HEASARC), provided by NASA's Goddard Space Flight Center. The authors thank Xian Hou and Jian Li for their helpful suggestions, and acknowledge support from the National Natural Science Foundation of China (Grant Nos. 12333007, 12373051) and the China Space Origins Exploration Program.

\bibliography{ref.bib}
\bibliographystyle{aasjournalv7}


\end{document}